# IrO$_2$ Surface Complexions Identified Through Machine Learning and Surface Investigations

*Jakob Timmermann[1], Florian Kraushofer[2], Nikolaus Resch[2], Peigang Li[3], Yu Wang[4], Zhiqiang Mao[3,4], Michele Riva[2], Yonghyuk Lee[1], Carsten Staacke[1], Michael Schmid[2], Christoph Scheurer[1], Gareth S. Parkinson[2], Ulrike Diebold[2], and Karsten Reuter[1,5,\*]*

[1]Chair for Theoretical Chemistry and Catalysis Research Center
Technical University of Munich, Lichtenbergstr. 4, D-85747 Garching (Germany)

[2]Institute of Applied Physics, Technical University of Vienna,
Wiedner Hauptstr. 8-10/134, A-1040 Vienna (Austria)

[3]Department of Physics and Engineering Physics, Tulane University,
New Orleans, LA 70118 (USA)

[4]Department of Physics, The Pennsylvania State University,
University Park, PA 16803 (USA)

[5]Fritz-Haber-Institut der Max-Planck-Gesellschaft,
Faradayweg 4-6, D-14195 Berlin (Germany)

\*corresponding author: karsten.reuter@ch.tum.de

Abstract: A Gaussian Approximation Potential (GAP) was trained using density-functional theory data to enable a global geometry optimization of low-index rutile IrO$_2$ facets through simulated annealing. *Ab initio* thermodynamics identifies (101) and (111) (1×1)-terminations competitive with (110) in reducing environments. Experiments on single crystals find that (101) facets dominate, and exhibit the theoretically predicted (1×1) periodicity and X-ray photoelectron spectroscopy (XPS) core level shifts. The obtained structures are analogous to the complexions discussed in the context of ceramic battery materials.

Keywords: Machine Learned Potentials, Rutile Oxides, Global Geometry Optimization, Complexions, Scanning Tunneling Microscopy, X-ray Photoelectron Spectroscopy

First-principles computations based on density-functional theory (DFT) have become a standard tool to determine surface structure. In the standard approach, a set of trial structures are optimized geometrically to identify minima on the ground-state potential energy surface (PES). Observables are computed to check for consistency with experimental data and one structure is declared best. While successful, this approach depends on the trial structures, and it is possible that the true surface is simply missed.

With the increasing efficiency of DFT calculations and computational power, DFT-based global geometry optimization has been heralded as a significant step to overcome this limitation.[1-6] Despite impressive successes of simulated annealing or basin hopping work, this direct approach has never truly affected the popularity of the 'trial set and local geometry optimization' approach. The excessive number of computations required by even the most efficient algorithms[7,8] leads to an intractable computational demand, particularly for reconstructions with large surface unit cells. Fortunately, machine-learned (ML) interatomic potentials[9,10] may now overcome this deadlock and enable a paradigm shift in our approach to automatic structure searches. These potentials can be trained with a feasible number of DFT calculations and, if needed, can be retrained on-the-fly in the course of an ongoing global geometry optimization. Crucially, the optimization is performed using the inexpensive ML potential, which enables extensive sampling of the configuration space.

Here, we use this approach to find the most stable surface terminations of rutile-structured oxides. Our motivation came from empirical reports that $IrO_2$ catalysts for proton-exchange membrane (PEM) water electrolysis exhibit increased activity following electrochemical activation with a small number of reductive formation cycles.[11,12] We hypothesized this might originate in a metal-rich complexion, which are discussed in the context of ceramic battery materials [13]. A complexion is a surface (or interfacial) phase that possesses a thermodynamically-determined equilibrium thickness on the order of nanometers, but is neither a thin version of a known 3D bulk phase, nor merely a reconstructed surface layer. While the ubiquity and importance of complex (often large surface unit-cell) reconstructions at surfaces of compound catalysts under operation conditions is well known [14-17], complexions can be more subtle by only involving deeper compositional changes at unchanged translational symmetry. After training a ML Gaussian Approximation Potential (GAP)[18,19] with DFT data, simulated-annealing-based global optimization immediately leads to very stable new terminations on the (101) and (111) low-index surfaces of rutile $IrO_2$ with mixed Ir-Ir and Ir-O bonding. Direct *ab initio* thermodynamics[20] calculations confirm the high stability of these complexions under strongly reducing conditions[21] – not only on $IrO_2$, but also on $RuO_2$, which is the alternative state-of-the-art rutile-structured catalyst used in PEM electrolysis. The theoretical predictions are supported by surface investigations of $IrO_2$ single crystals, which exhibit (101) facets rather than the more common low-energy (110) orientation of rutile.[22,23] Characterization by low-energy electron diffraction (LEED), scanning-tunneling microscopy (STM) and X-ray photoelectron spectroscopy (XPS) confirms the properties of the predicted metal-rich complexions, explaining why $IrO_2$ nanoparticles often expose (101) facets.[24-30]

Our investigation starts with the creation of a reference database of DFT structures to train the non-parametric GAP potential. GAPs decompose the total energy of a system into a sum of atomic energies that depend on the local chemical environment.[18,19] This dependence is learned from the atomic environments present in the reference database through Gaussian process regression. For energy predictions, the similarity between each atom in an unknown structure and representative training atoms is then determined via a kernel function. In this work, we employ the smooth overlap of atomic positions (SOAP) kernel,[31] which considers all neighboring atoms within a radius of 5.5 Å, combined with a simple two-body kernel based on interatomic distances. The reference database comprises 136 structures calculated with QuantumEspresso[32] and the RPBE[33] exchange-correlation functional. These structures span a range of most diverse chemical environments, and comprise various optimized or near-optimum crystalline bulk and low-index surface geometries of different stoichiometry, as well as highly non-equilibrium structures taken from snapshots of high-temperature molecular dynamics (MD) simulations of differently shaped and sized nanoparticles. Validated against an equally diverse set of 39 structures not used in the training, the final GAP reproduces the widely varying DFT formation energies with a mean average error of 25 meV/atom.

To explore a possible formation of complexions, we performed extensive simulated annealing MD runs for all five symmetry-inequivalent low-index surfaces[34] of rutile $IrO_2$, each time starting with the metal-rich regular (1×1) termination expected in reducing conditions. Specifically, we employ periodic boundary condition supercells with thick slabs comprising at least seven rutile trilayers and (3×3) or (4×4) surface unit cells as further detailed in the SI. The temperature is initially raised to around 1000 K for 20 ps, before a slow cooling rate of 3 K/ps is applied during an additional 250 ps. After a final geometry optimization, we obtain new structures with a significantly lower energy, in particular for the (101) [or the symmetry-equivalent (011)] and the (111) orientations. Analysis of these structures (**Figure 1**) reveals that neither correspond to a reconstruction with a lowered translational symmetry, but is instead a reordering of the original rutile layering sequence that preserves the regular (1×1) lateral periodicity. Direct recalculation and geometry re-optimization of these structures at the DFT RPBE level confirms the reliability of the GAP prediction. The structures are significantly lower in energy than the regular Ir-rich (1×1) termination for the respective orientation and, in slabs where at least the five topmost layers are allowed to move, the regular Ir-rich (101) termination relaxes barrierlessly into the new complexion.

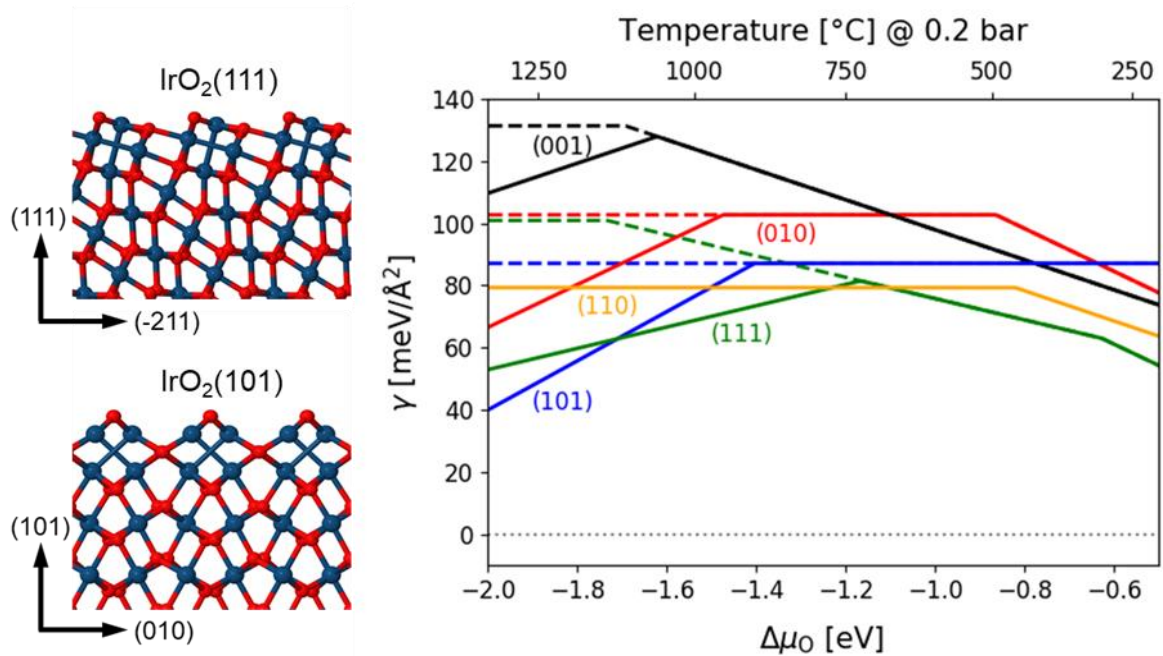

**Figure 1:** (Left) Side views of identified complexions. Ir and O atoms are drawn as blue (large) and red (small) spheres, respectively. (Right) Computed surface free energies $\gamma$ of the five symmetry-inequivalent low-index facets in a pure oxygen atmosphere. In the top $x$ axis, the dependence on the oxygen chemical potential $\Delta\mu_O$ is translated into a temperature scale at 0.2 bar pressure (the oxygen partial pressure in air). The dashed lines indicate the surface free energies without complexions.

Within an *ab initio* thermodynamics framework[20] we can compare the stability of these new structures to all other possible and known (1×1) terminations of rutile $IrO_2$. Generally, there is at least one metal-rich, one stoichiometric and one oxygen-rich termination for each low-index facet, with some facets lacking some terminations and some facets additionally allowing for an oxygen-superrich termination (Figs. S7-S11).[35] Figure 1 shows the resulting surface phase diagram. Only the lowest surface free energy is shown for each facet as a function of the chemical potential of oxygen $\Delta\mu_O$, and a kink in the surface free energy line reflects a change in the most stable termination. Metal-rich terminations exhibit a positive slope, O-rich terminations a negative slope, and stoichiometric terminations are independent of $\Delta\mu_O$. For low oxygen chemical potentials, the complexions significantly lower the surface free energy and change the relative energetic ordering. The same form and relative ordering of the surface free energies is obtained with the stronger-binding PBE[36] functional, but the entire phase diagram shifts to lower $\Delta\mu_O$ (see Fig. S13). While we cannot quantify the gas-phase conditions of the phase stability, the (110) facet would clearly be the lowest-energy orientation in reducing environments in the absence of complexations, in line with the predominant focus of surface science work on this particular facet.[22,23] The stability of the complexions makes the (101) and (111) facets energetically competitive.

In order to test this surprising finding, we investigate the surfaces of IrO$_2$ single crystals grown in a tube furnace with an O$_2$ inflow of 100 ml/min at atmospheric pressure. Ir was supplied from Ir powder (99.99%) at 1250 °C and flake-shaped IrO$_2$ crystals formed at the colder end of the furnace (1000 °C). Two of the larger crystallites (both ca. 3 mm$^2$ top surface area) were chosen for surface studies. Electron backscatter diffraction (EBSD) immediately reveals that all areas where a diffraction pattern could be identified expose (101)-type surfaces (Fig. S18).

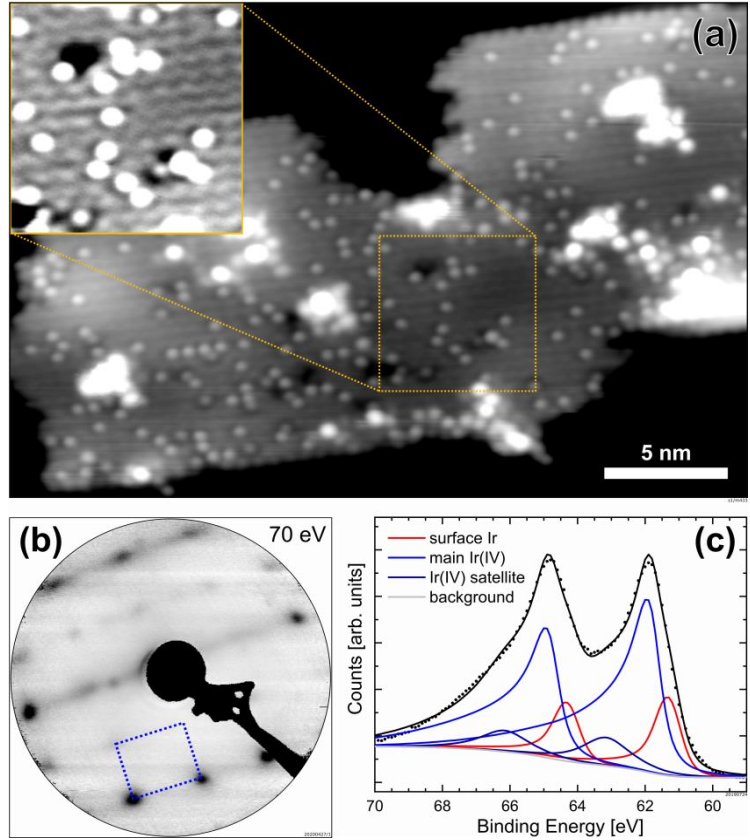

**Figure 2:** IrO$_2$ crystal after UHV preparation. (a) STM image of the IrO$_2$ sample (30 × 20 nm$^2$, $U_{\text{sample}} = -2$ V, $I_{\text{tunnel}} = 0.2$ nA) with the inset processed to enhance the visibility of the atomic corrugation. (b) LEED pattern from one of the IrO$_2$ crystals ($E_{\text{electron}} = 70$ eV). A rectangular pattern (unit cell drawn in blue) is clearly visible, and its diffraction spots move towards the center of the screen with increasing energy, as expected for normal beam incidence. (c) XPS data (points) and fit (lines) of the Ir 4$f$ region (Mg Kα anode).

The two IrO$_2$ samples were then loaded to an ultra-high vacuum (UHV) chamber and prepared by sputtering (1 keV Ar$^+$ ions, $p_{\text{Ar}} = 8 \times 10^{-6}$ mbar, 10 min) and annealing in oxygen (450 °C, 40 min). Oxygen gas was dosed through an oxygen shower, with the gas outlet very close to the sample. This increases the local pressure by a factor of ≈10–20 compared to the O$_2$ pressure measured in the chamber ($5 \times 10^{-6}$ mbar).

Low energy electron diffraction (LEED) images (**Figure 2**b and Figure S20) reveal a rectangular pattern with the spots moving towards the center of the screen with increasing incident beam energy. The unit cell size of $(0.58 \pm 0.04) \times (0.47 \pm 0.04)$ nm$^2$ was quantified using LEED patterns of a Pd(111) single crystal as a reference. These numbers are in good agreement with the (1×1) unit cell of IrO$_2$(101) ($0.55 \times 0.45$ nm$^2$). Some additional diffraction spots are also observed, but these move in non-radial directions with increasing energy, indicating the presence of other facets for which the beam incidence is off-normal.

STM images acquired at room temperature after UHV preparation (**Figure 2**a) exhibit a zig-zag pattern with (1×1) symmetry. Since the intrinsic drift of the STM scanner cannot be corrected by comparison to a known structure, distance measurements are unreliable, but the $0.55 \times 0.45$ nm$^2$ spacing expected for an IrO$_2$(101)-(1×1) unit cell fits the data within the expected error. Some bright point features are also visible, which we attribute to either lattice defects or adsorbates. Interestingly, the periodicity of these protrusions cannot be reconciled with a bulk-truncated (1×1) surface (Figure S22). On the proposed complexion, the features are located at surface oxygen sites, allowing tentative assignment as either surface hydroxyls or oxygen vacancies.

On one of the samples, a second, pseudo-hexagonal surface phase was also observed (Fig. S19). The nearest-neighbor distances were determined as ≈0.55 nm, which would fit an IrO$_2$(111)-(1×1) unit cell. However, since the unit cell angle cannot be accurately determined by STM alone, a (2×2) superstructure on Ir(111) could, in principle, also fit the data. We also note that a hexagonal reconstruction of the (101) facet was previously observed on rutile TiO$_2$(101),[37] and attributed to contamination.

XPS of the Ir 4*f* region is shown in **Figure 2**c. We fit the spectrum using peak shapes and oxide satellite peak positions from ref. [38], which results in a peak at 61.9 eV (blue) due to Ir(IV), and a lower-binding-energy component shifted by 0.6 eV at 61.3 eV (red). This agrees with our initial-state calculations of the 4*f* DFT Kohn-Sham orbital positions for the $IrO_2$(101) complexion, which yield an initial-state shift of 0.6 eV for the top two Ir layers with respect to a bulk-like Ir using both the RPBE and PBE functionals (see SI). A much larger shift (1.1 eV towards lower binding energies) is predicted for the top layer of the regular Ir-rich (101) termination. We also acquired Ir 4*f* peak data from a freshly sputtered sample (Fig. S21), which is dominated by a strong contribution of metallic surface iridium at 60.9 eV, in agreement with the position reported in the literature for Ir single crystals.[31] Overall, the experimental evidence clearly shows that the crystals are dominated by (101) facets with a (1×1) surface symmetry, and support the predicted complexed $IrO_2$(101)-(1×1) surface. Since the crystal growth direction was not enforced and the relatively rough, vicinal surfaces would have faceted to a more stable orientation, the dominance of (101) surface is an indirect confirmation of its thermodynamic stability at the growth conditions.

In our view, the complexed surfaces are precursors to a full reduction of the bulk oxide. Two layers with mixed Ir–O and Ir–Ir bonding are achieved through a mere reordering of the rutile layering sequence the (101) and (111) orientations, and the increased coordination of the topmost Ir atoms (from 3-fold to 4-fold) stabilizes the structures. Adding further equivalent complexion layers does not further increase this coordination, and we calculate higher surface free energies for such structures (see Fig. S12). As such, the identified complexions are novel 2D interphases and not just thin versions of known 3D bulk structures, and are thus quite analogous to the much discussed surface oxides as precursors to a full oxidation of transition metals.[39-41]

With this understanding, one would expect complexions to be a general feature of oxides in reducing environments, and follow up computations predict that analogous complexions render the (101) and (111) facets energetically competitive for rutile $RuO_2$ under reducing conditions (Figure S14).

In summary, a completely unexpected class of surface structures was readily identified for a well-studied type of oxide crystals using ML interatomic potentials. That such simple structures have consistently eluded previous trial-structure-based surface structure determination work on $IrO_2$ or $RuO_2$ shows them to be counterintuitive, and one wonders how many more surprises await us when global geometry optimization based on predictive-quality machine-learned potentials has reached full maturity.

Supporting Information Available

See Supporting Information [42] for additional detailed information on the reference database employed for the training of the ML potential and simulated annealing runs, on the *ab initio* thermodynamics results, on the initial-state core level shift calculations, as well as on the experimental growth and characterization work. This material is available free of charge via the Internet at http://pubs.acs.org. All in- and output files for the DFT training structures are available at the NOMAD database and can be accessed under DOI: 10.17172/NOMAD/2020.08.23-2 [1].

ACKNOWLEDGEMENT

This research was supported by the Kopernikus/P2X programme (Cluster FC-A1) of the German Federal Ministry of Education and Research, the German Federal Environmental Foundation DBU and the German Academic Exchange Service DAAD. UD, MR and FK acknowledge support by the Austrian Science Fund (FWF, Z-250, Wittgenstein Prize). GSP acknowledges funding from the European Research Council (ERC) under the European Union's Horizon 2020 research and innovation program Grant agreement No. 864628. NR was supported by the Austrian Science Fund (FWF, Y847-N20, START Prize). ZQM acknowledges the support from the US National Science Foundation under grant DMR 1917579. We acknowledge fruitful discussions with Johannes Margraf and Simon Wengert. The authors thank Andreas Steiger-Thirsfeld (USTEM, TU Wien) for support with SEM measurements.

TOC Graphics:

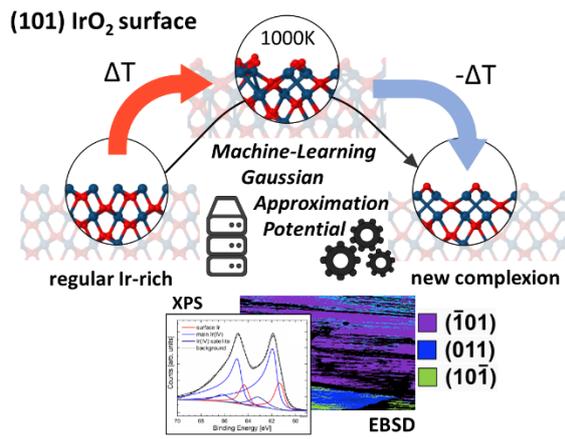